\documentclass[preprints,article,accept,moreauthors,pdftex]{Definitions/mdpi}

\firstpage{1} 
\pubvolume{1}
\issuenum{1}
\articlenumber{0}
\pubyear{2023}
\copyrightyear{2023}
\datepublished{ } 
\hreflink{https://doi.org/} 

\pdfoutput=1

\usepackage[separate-uncertainty=true]{siunitx}

\Title{ARIADNE$^{+}$: Large Scale Demonstration of Fast Optical Readout for Dual-Phase LArTPCs at the CERN Neutrino Platform
$^{\dagger}$}
\TitleCitation{ARIADNE$^{+}$: Large Scale Demonstration of Fast Optical Readout for Dual-Phase LArTPCs at the CERN Neutrino Platform}

\Author{Adam John Lowe\orcidA{} 
 $^{1,}$* 
, 
 Pablo Amedo-Martinez $^{2}$, Diego González-Díaz $^{2}$, Alexander Deisting $^{3}$, Krishanu Majumdar $^{1}$, Konstantinos~Mavrokoridis $^{1}$, Marzio Nessi $^{4}$, Barney Philippou $^{1}$, Francesco Pietropaolo $^{4}$, Sudikshan~Ravinthiran~$^{1}$, Filippo~Resnati $^{4}$, Adam Roberts $^{1}$, Angela Saá Hernández $^{2}$, Christos Touramanis $^{1}$ and~Jared~Vann $^{1}$}

\AuthorNames{Adam Lowe, Pablo Amedo, Diego González-Díaz, Alexander Deisting, Krishanu Majumdar, Konstantinos Mavrokoridis, Marzio Nessi, Barney Philippou, Francesco Pietropaolo, Sudikshan Ravinthiran, Filippo Resnati, Adam Roberts, Angela Saá Hernández, Christos Touramanis and Jared Vann}
\AuthorCitation{Lowe, A.; Amedo, P.; González-Díaz, D.; Deisting, A.; Majumdar, K.; Mavrokoridis, K.; Nessi, M.; Philippou, B.; Pietropaolo, F.; Ravinthiran, S.; et al.}
\address{%
$^{1}$ \quad Department 
 of Physics, University of Liverpool, Oliver Lodge Bld, Oxford Street, Liverpool L69 7ZE, UK;  k.majumdar@liverpool.ac.uk (K.M.); kostasm@liverpool.ac.uk (K.M.); (barney.philippou@gmail.com (B.P.);  s.ravinthiran@liverpool.ac.uk (S.R.); a.roberts7@liverpool.ac.uk (A.R.); c.touramanis@liverpool.ac.uk (C.T.); jared@vann.me (J.V.) 
\\
$^{2}$ \quad Instituto Galego de Física de Altas Enerxías, Rúa de Xoaquín Díaz de Rábago, s/n,Campus Vida, Universidade de Santiago de Compostela, 15705 Santiago de Compostela 
 Spain; pablo.amedo.martinez@usc.es (P.A.); diego.gonzalez.diaz@usc.es (D.G.-D.); angela.saa.hernandez@usc.es~(A.S.H.)\\
$^{3}$ \quad Department of Physics, Royal Holloway University of London, Tolansky Building, Egham, \mbox{Surrey TW20 0EX, UK}; alexander.deisting@cern.ch\\
$^{4}$ \quad European Organization for Particle Physics (CERN),  P.O. Box
1211, Geneva
, Switzerland; marzio.nessi@cern.ch (M.N.); francesco.pietropaolo@cern.ch (F.P.); filippo.resnati@cern.ch (F.R.)\\

}

\corres{Correspondence: a.j.lowe@liverpool.ac.uk}

\conference{NuFACT 2022.} 

\abstract{Optical readout of large scale dual-phase liquid Argon TPCs is an attractive alternative to charge readout and has been successfully demonstrated on a 2~$\times$~2 m active region within the CERN protoDUNE cold box. ARIADNE$^{+}$ uses four Timepix3 cameras imaging the S2 light produced by 16 novel, patent pending, glass THGEMs. ARIADNE$^{+}$ takes advantage of the raw Timepix3 data coming natively 3D and zero suppressed with a 1.6 ns timing resolution. Three of the four THGEM quadrants implement readouts in the visible light range through wavelength shifting, with the fourth featuring a VUV light intensifier, thus removing the need for wavelength shifting altogether. Cosmic ray reconstruction and energy calibration were performed. Presented is a summary of the detector setup and experimental run, preliminary analysis of the run data and future outlook for the ARIADNE program.}


\begin{document}
\setcounter{secnumdepth}{2}
\noindent 
\begin{center}
\includegraphics[width=0.60\textwidth]{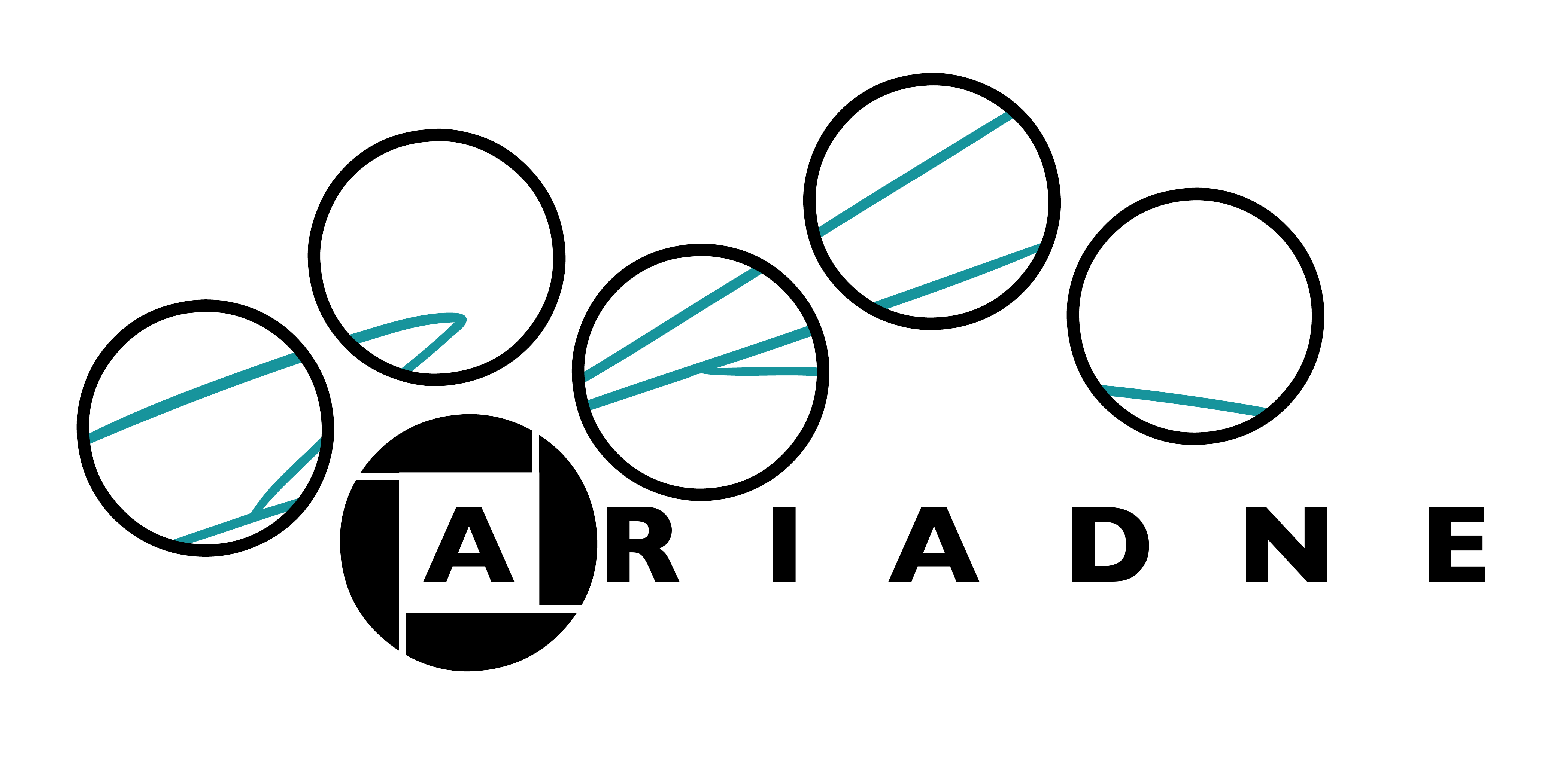}
\end{center}
\section{Introduction} 

With the size of Liquid Argon Time Projection Chambers (LArTPCs) ever increasing, and the cost of constructing and operating such detectors following a similar trend, the importance of R\&D into alternative readout methods has never been greater. The ARIADNE (AR
gon ImAging DetectioN chambEr) Experiment, a 1-ton dual-phase LArTPC, has demonstrated the viability of optical readout as an alternative to charge readout \cite{ARIADNE_TDR}, and ARIADNE$^{+}$ tests the feasibility of scaling up this technology further.

The ARIADNE program utilises the 1.6 ns timing resolution and native 3D raw data of the Timepix3 camera to image the wavelength-shifted secondary scintillation light generated by a THGEM (THick Gaseous Electron Multiplier) within the gas phase of the dual-phase LArTPC. The main detection principle sees incoming particles ionising LAr and creating prompt scintillation light (known as S1). The ionisation electrons then drift towards an extraction grid situated below the liquid level where they are transferred to the gas phase and subsequently amplified using a THGEM. The drift charge multiplication produces secondary scintillation light (S2), which is wavelength-shifted before imaging with a Timepix3 camera. 

With no need for thousands of internal charge TPC readout channels, pre-amps, etc., reduction in construction costs is one of a number of advantages to ARIADNE technology. The combination of an objective lens and Timepix3 specifications (excellent tracking capabilities, high timing and energy resolution and data suppression) enables the imaging of large tracking detector planes, which provides a very cost effective commercial system, and this is detailed further in \cite{ARIADNE_TPX}.   

One THGEM-accelerated electron can generate 100 s of scintillation photons, increasing sensitivity to low energies; Timepix3 is sensitive to single photons providing a high signal-to-noise ratio. The cameras are mounted externally relative to the TPC, this means low noise (decoupled from TPC electronics) and ease to swap out and repair technology.  

The move within ARIADNE from EMCCDs (Electron Multiplying CCDs) to Timepix3 brought with it natively 3D readout. The Timepix3 data stream consists of 64-bit ‘hit’ packets, and these packets contain the column and row number of the hit pixel (the XY of the hit), the Time of Arrival (the absolute Z position if the electron drift velocity is known) and Time over Threshold (analogous with intensity). Given the pixel-driven readout, as compared to frame-based readout, events come with zero background suppression and therefore require only a few kilobytes of storage.

The protoDUNE cold box located at the CERN Neutrino Platform offers a 5
~$\times$~5 m cryogenic vessel with a cathode as a test bed for scaling up ARIADNE technology with the support and expertise of the Neutrino Platform team. This is a key step in validating the feasibility of ARIADNE readout for kilo-tonne LAr detectors such as the DUNE far~detector. 

\section{The ARIADNE\boldmath{$^{+}$} Detector}

\begin{figure}[H]
   \hspace{-3.5em} \includegraphics[width=0.9\textwidth,height=75mm]{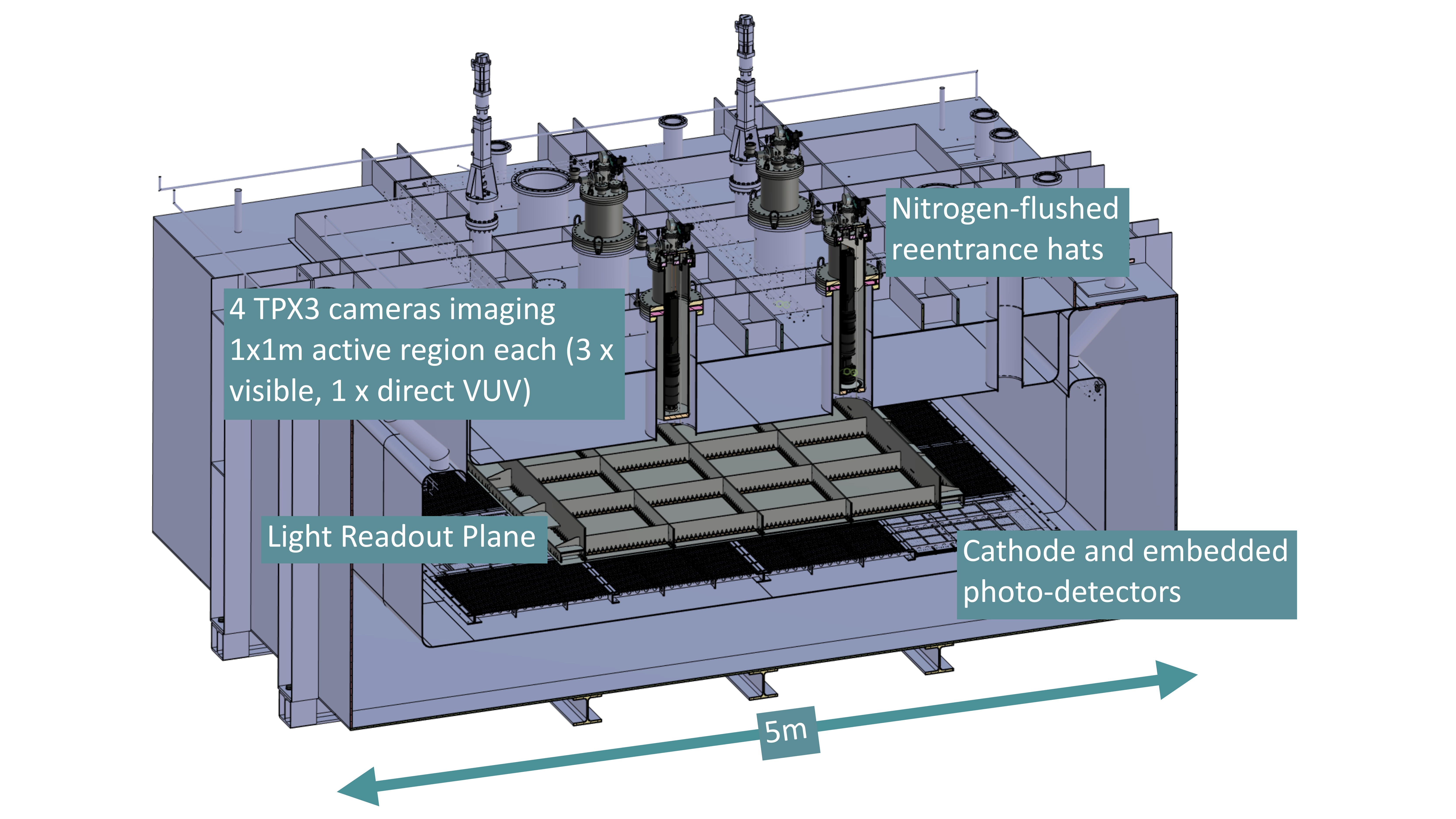}
    \caption{The 
 cold box at the Neutrino Platform in optical configuration.}
    \label{fig:cross-section of box}
\end{figure}

Testing optical readout on a scale relevant for DUNE, the ARIADNE$^{+}$ detector was built within the protoDUNE cold box located at the CERN Neutrino Platform and had an experimental run from February to April of 2022. The cold box itself is a 15-tonne cryogenic vessel, refurbished in 2021 to accommodate testing of vertical drift CRPs (Charge Readout Planes for DUNE). For ARIADNE$^{+}$, a LRP (Light Readout Plane) was mounted underneath the lid of the cold box, suspended 20 cm above the cathode and using four Timepix3 cameras to image an active area of 2~$\times$~2 m, as shown in Figure \ref{fig:cross-section of box}.  

\subsection*{LRP (Light Readout Plane)}
\begin{figure}[H]
    \includegraphics[width=0.54\textwidth,height=43mm]{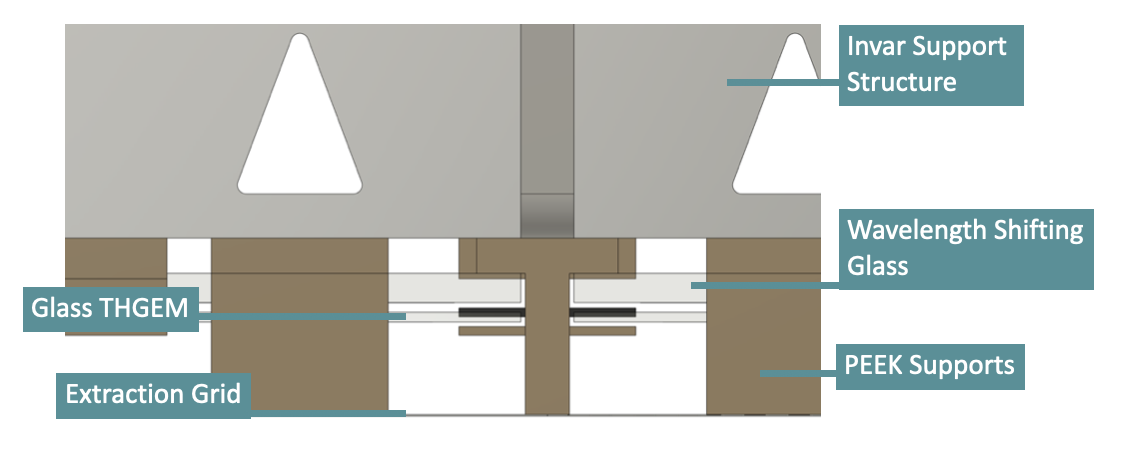}
        \includegraphics[width=0.39\textwidth,height=45mm]{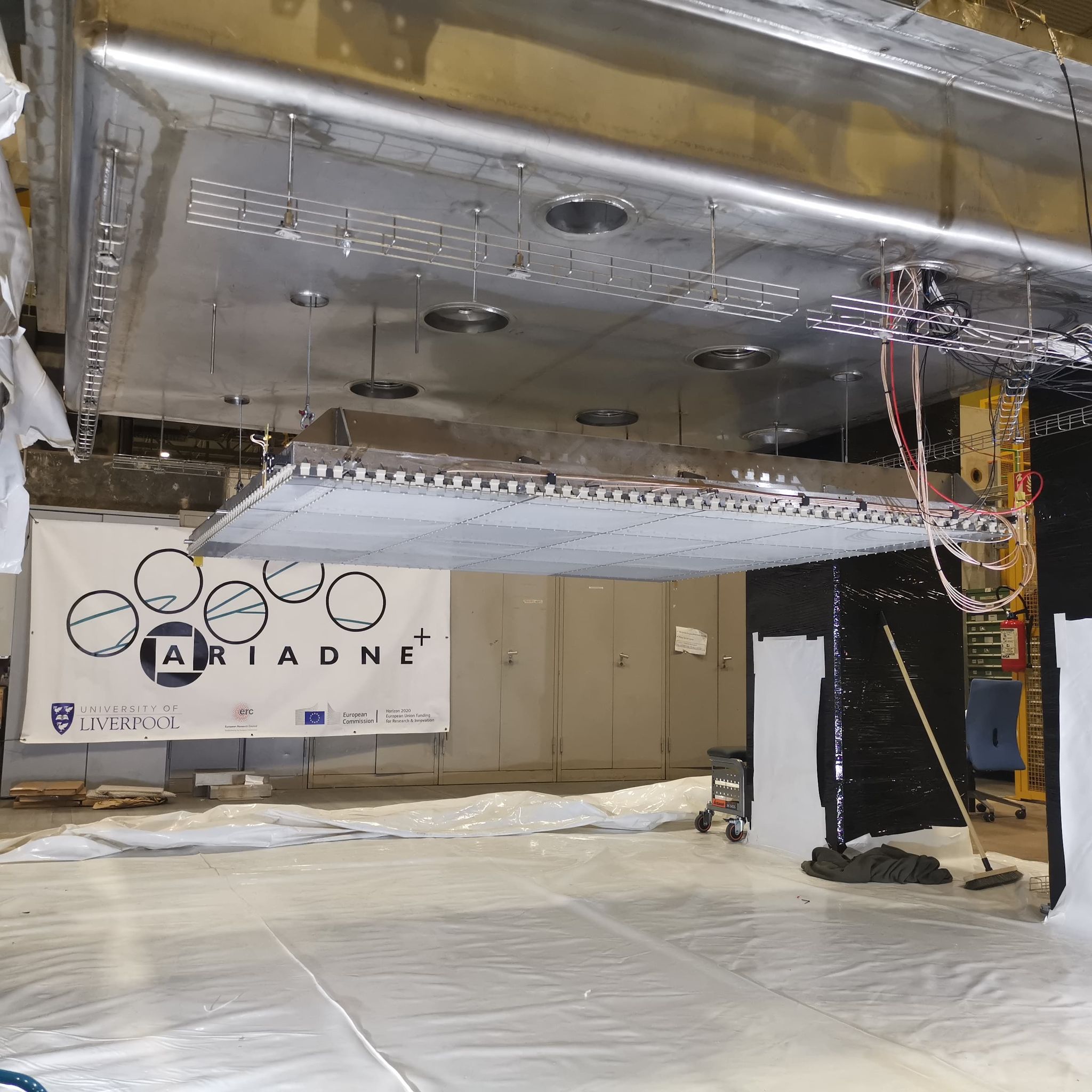}
    \caption{Cross-section 
 of the ARIADNE$^{+}$ LRP assembly (\textbf{Left}) and LRP integration (\textbf{Right}).}
    \label{fig:cross-section}
\end{figure}

The LRP itself is comprised of a 2.3~$\times$~2.3 m Invar support frame, chosen for its uniquely low coefficient of thermal contraction, and an assembly of an extraction grid, THGEM and WLS PEN glass (Figure \ref{fig:cross-section}). The extraction grid is made up of chemically-etched stainless steel pieces of a modular design, intended to reduce sag across the area of the LRP. The extraction grid is mounted 15 mm from the 50~$\times$~50 cm glass THGEMs. Liverpool University patent approved glass THGEMs \cite{GLASS_THGEMS} have increased rigidity over FR4 THGEMs, vital for area sizes such as those within ARIADNE$^{+}$, and feature `hour-glass'-shaped holes, which collect charge over time and increase light output at lower biases than FR4 THGEMs. For the three 1~$\times$~1 m quadrants with Timepix3 cameras imaging visible light, situated above the THGEMs is a PEN film-coated glass. 

The Timepix3 camera images direct VUV (Vacuum Ultraviolet) light via a VUV intensifier. A custom-made 5 mm diameter Magnesium-Fluoride lens focuses light from a 1~$\times$~1~m active area onto the intensifier's photo-cathode for imaging with the Timepix3~camera. 

 For the duration of the three week ARIADNE$^{+}$ run, the purity was continuously monitored by the CERN Neutrino Platform team, and an electron lifetime of approximately 0.5~ms 
 was ensured throughout. S1 data were also collected using X-ARAPUCAS embedded within the cold box cathode for S1/S2 analysis, which is ongoing. 

\section{Results}
\subsection{Gallery of Events}
\begin{figure}[H]
    \includegraphics[width=0.45\textwidth,height=47mm]{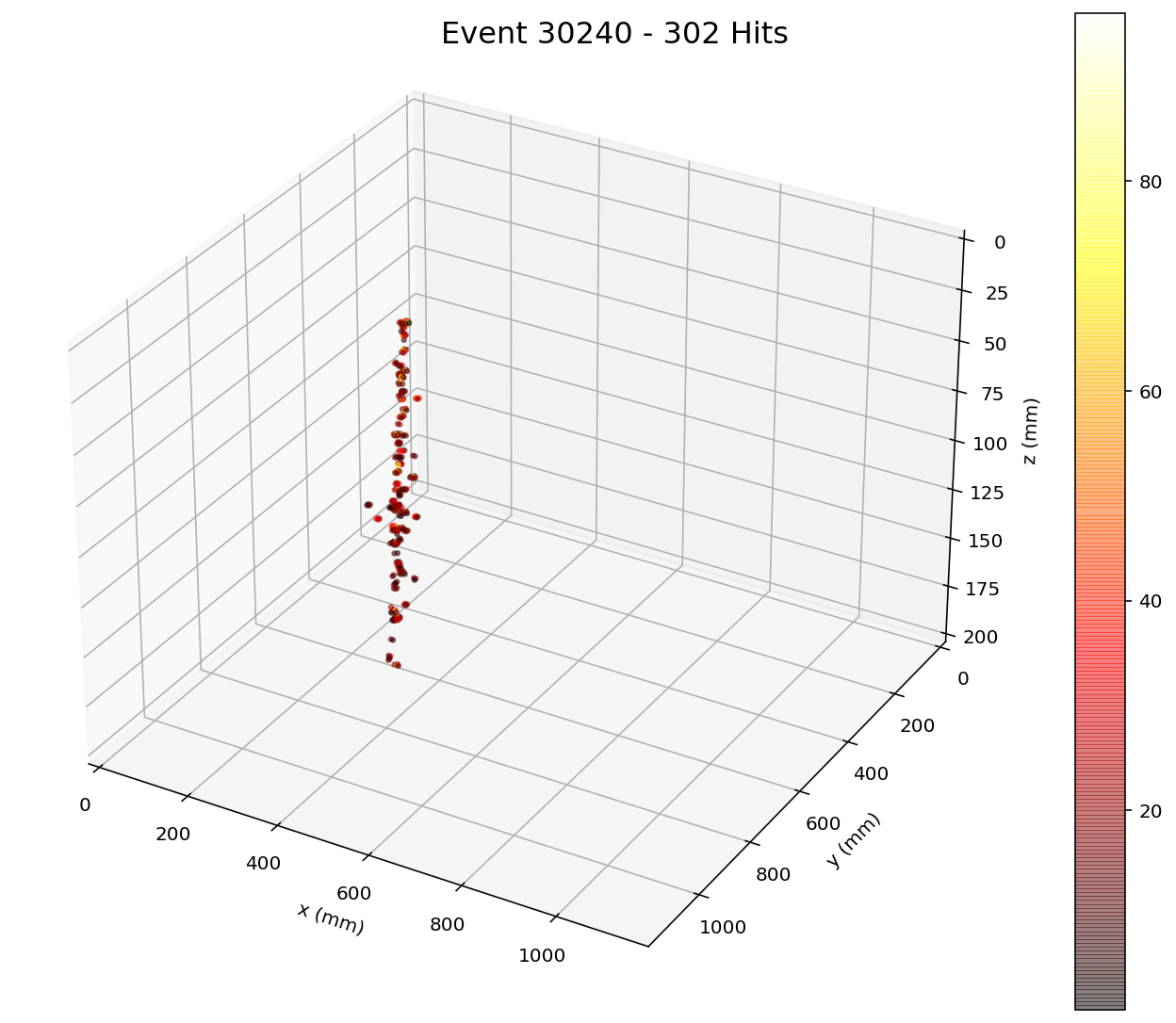}
    \includegraphics[width=0.45 \textwidth,height=47mm]{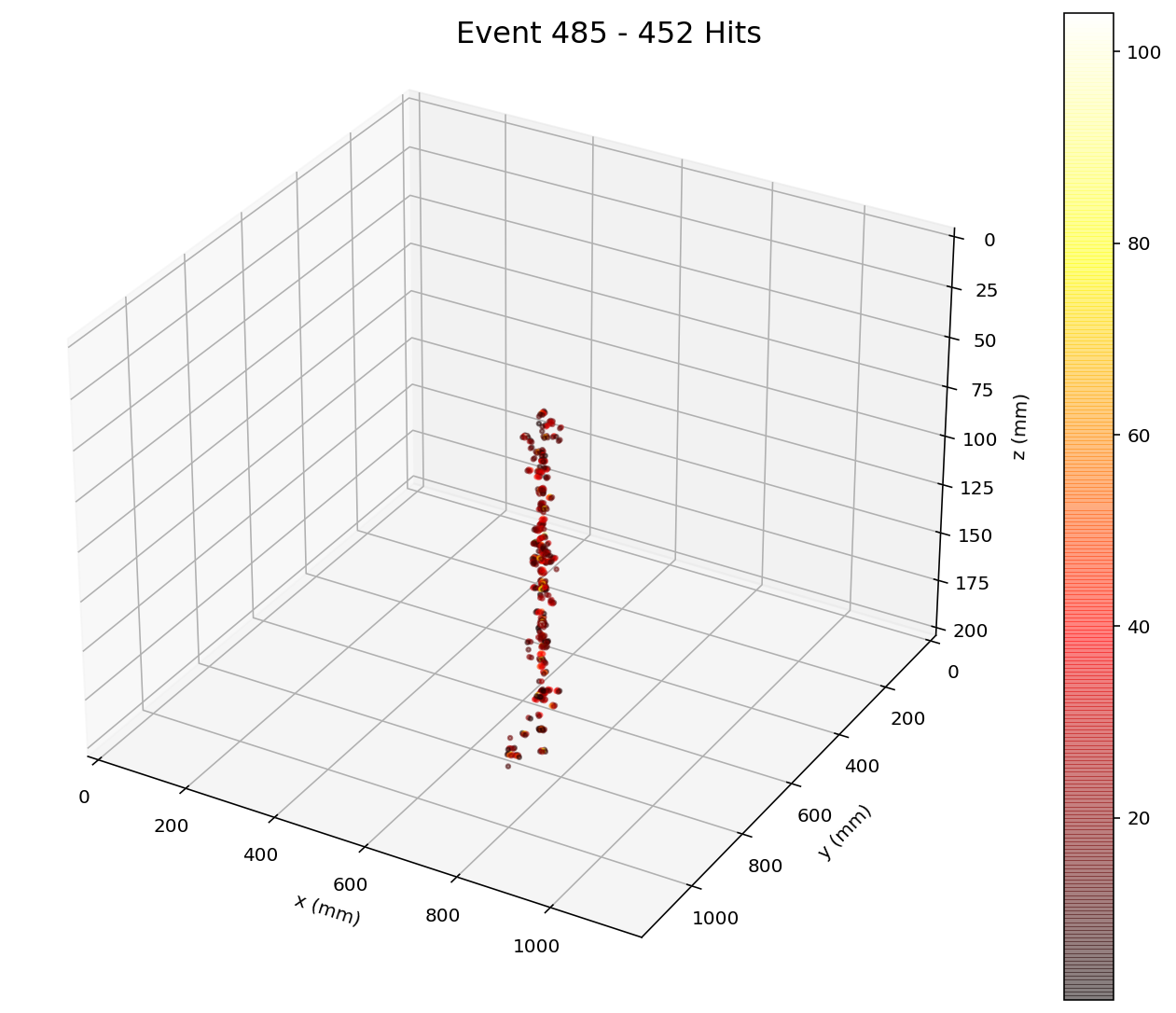}
    \caption{Visible events with a spatial resolution of approx. 4 mm.}
    \label{fig:visible_events}
\end{figure}

\begin{figure}[H]
    \includegraphics[width=0.45\textwidth,height=47mm]{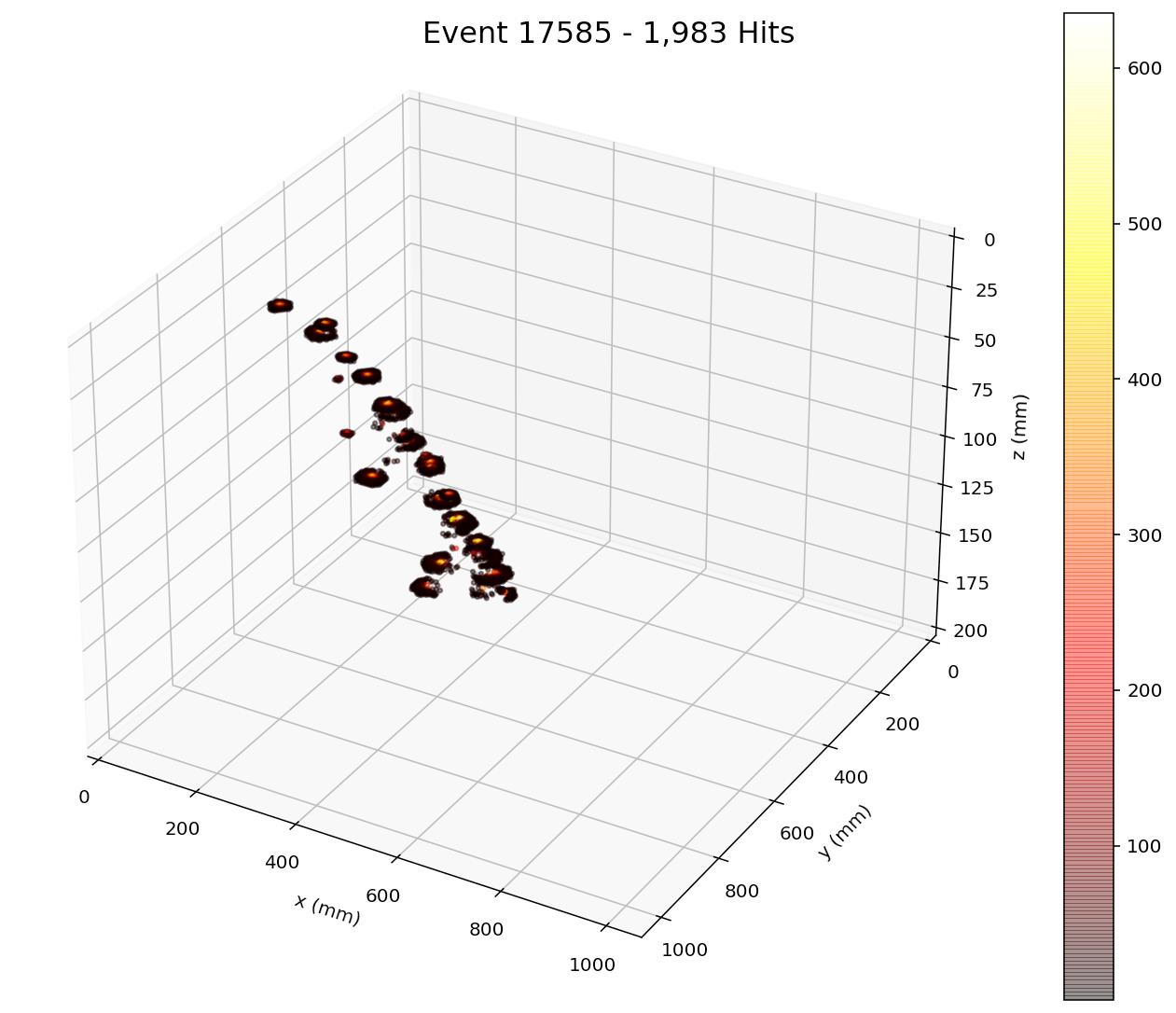}
    \includegraphics[width=0.45\textwidth,height=47mm]{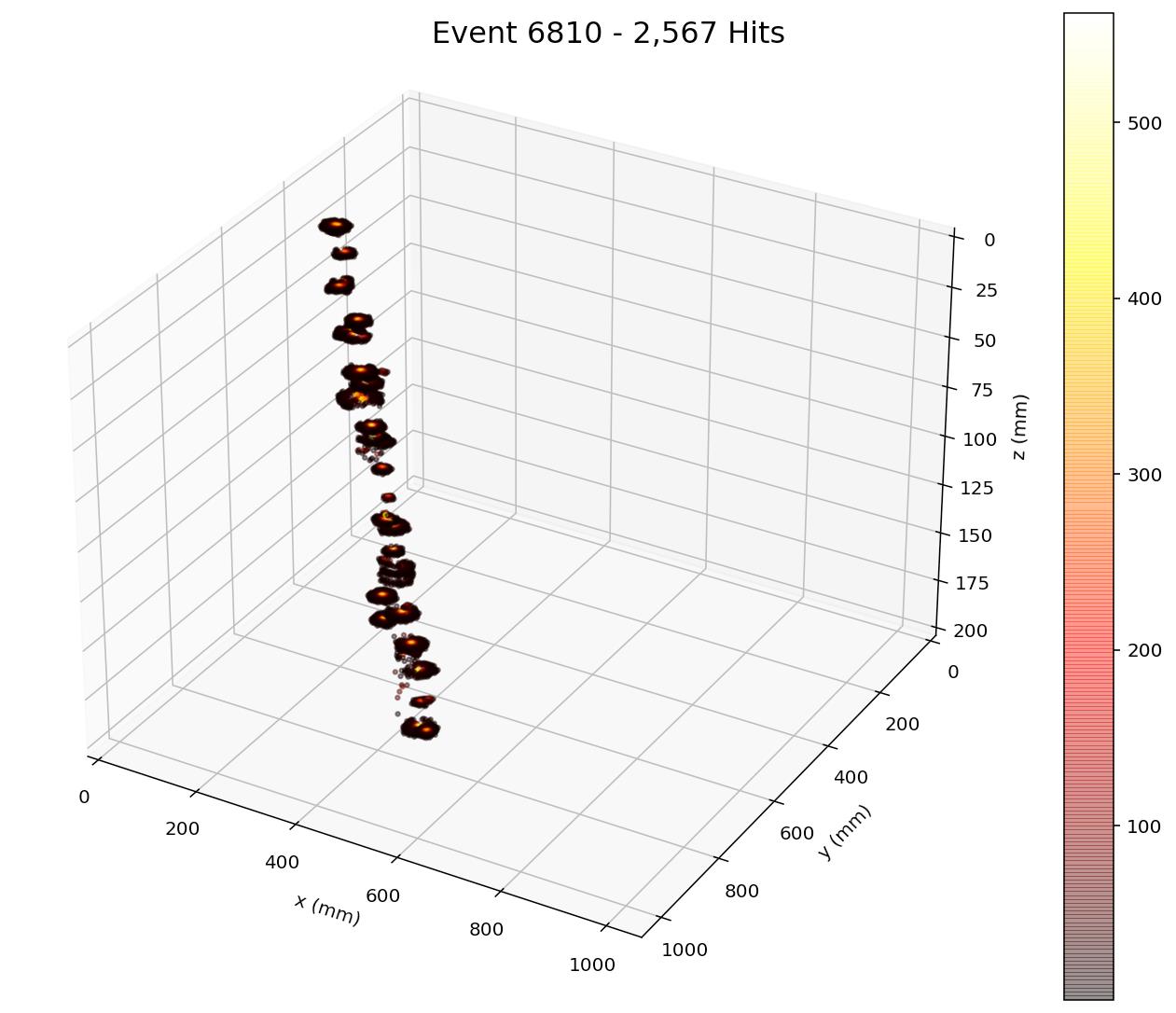}
    \caption{VUV 
 events with a spatial resolution of approx. 4 mm.}
    \label{fig:VUV_events}
\end{figure}

Three weeks of cosmic data were taken with the ARIADNE$^{+}$ detector, and Figure \ref{fig:visible_events} is a selection of through-going muons, from one camera imaging visible light over 1~$\times$~1 m active area, with a spatial resolution of approximately 4 mm. Figure \ref{fig:VUV_events} is a selection of, again, through-going muons, imaged this time using a VUV intensifier over \hl{1}~$\times$~1 m, with again approximately 4 mm spatial resolution. 

\subsection{30 s Exposure Cosmics}

\begin{figure}[H]
    \includegraphics[width=0.38\textwidth,height=45mm]{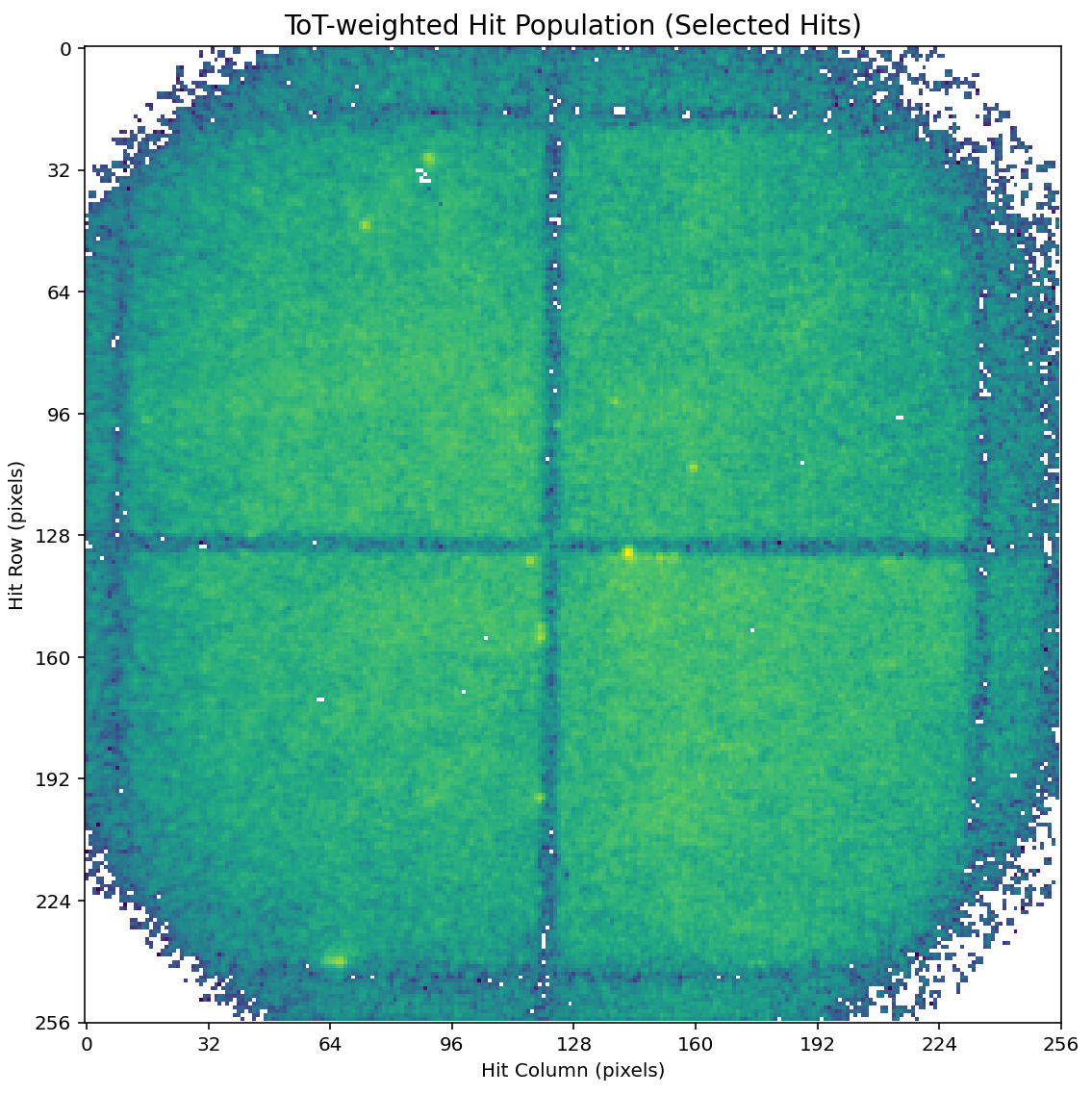}
    \includegraphics[width=0.38\textwidth,height=46mm]{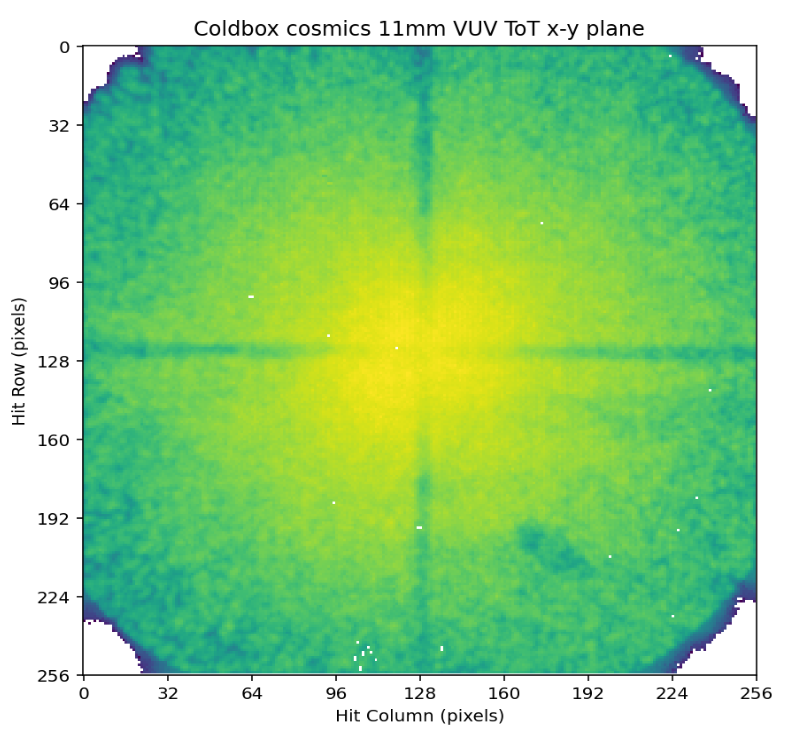}
    \caption{A 30 s cosmic ray exposure with visible light (\textbf{Left}) and VUV light (\textbf{Right}).}
    \label{fig:exposure}
\end{figure}

When the total ToT is summed for 30 s for both visible light and VUV, Figure \ref{fig:exposure} is produced. This is for one camera imaging 1~$\times$~1 m active area, i.e., one quadrant of the LRP, comprising of four THGEMs. The VUV light was focused using a custom MgF$_{2}$ lens, which does have noticeable vignetting effects; this will be corrected in future analysis or further development of VUV optics.       
\vspace{-6pt}

\subsection{Energy Calibration and Resolution}

\begin{figure}[H]
    \includegraphics[width=0.48\textwidth,height=45mm]{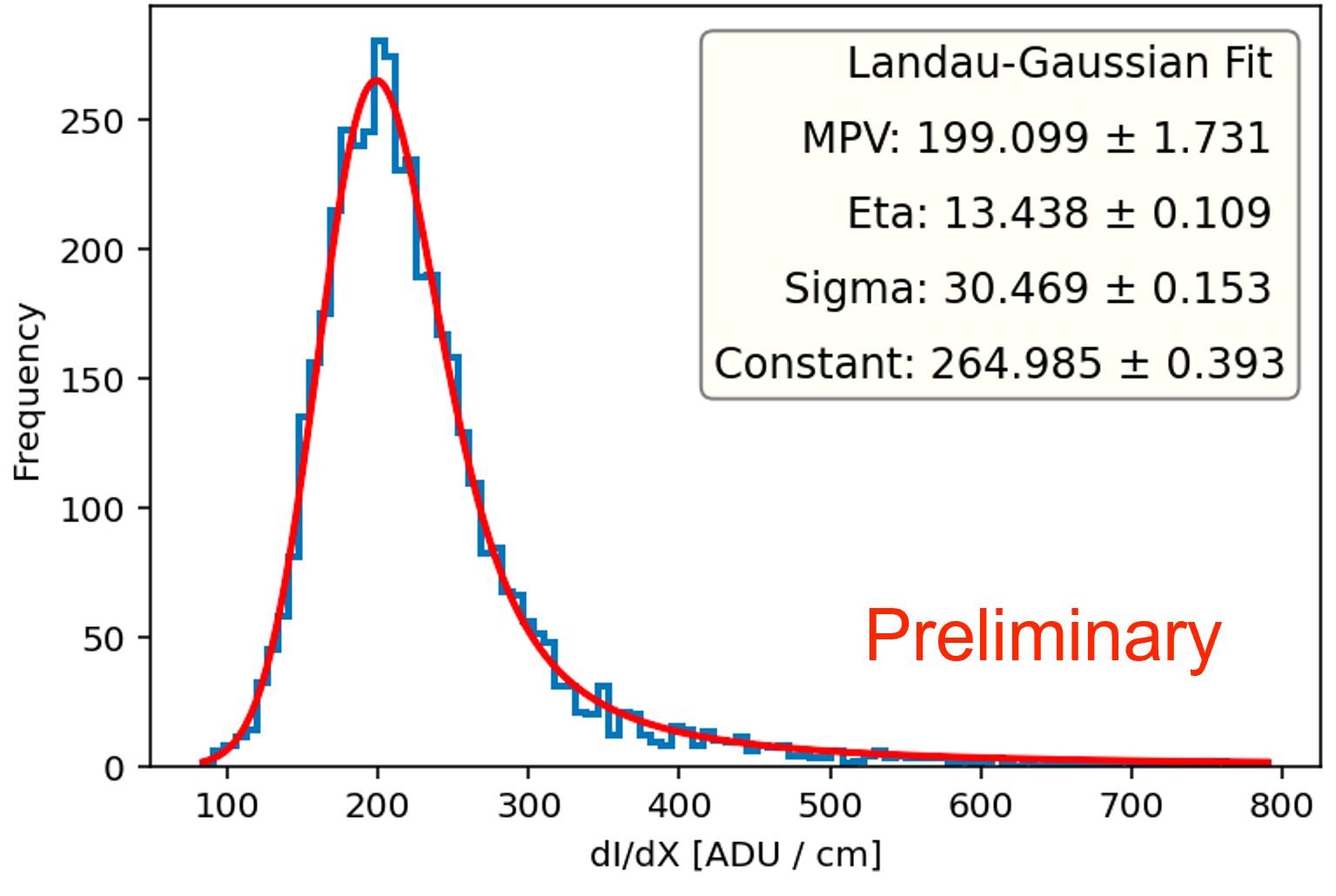}
    \includegraphics[width=0.54\textwidth,height=45mm]{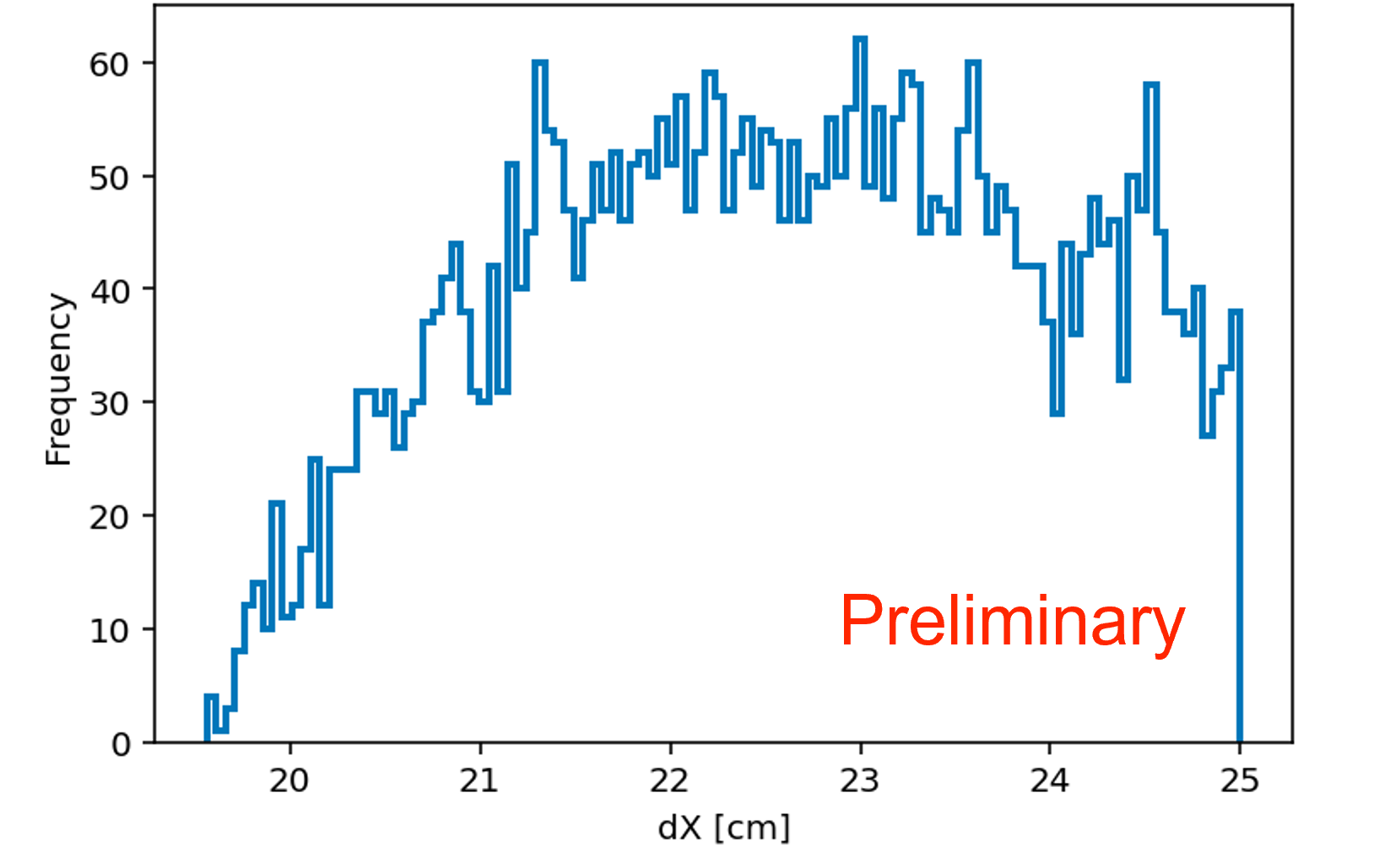}
    \caption{The 
 dI/dX distribution (\textbf{Left}) and the distribution of events used for calculating the calibration and energy resolution (\textbf{Right}).}
    \label{fig:resolution}
\end{figure}

By selecting only through-going muons after track-fitting (i.e., muons that go through the THGEM and greater than 19 cm in depth), it is possible to obtain an energy calibration and resolution given muons' well-known mean energy deposition rate in LAr of 2.12~MeV/cm \cite{mipCalibration}. The process for obtaining these values is given in greater detail in \cite{ARIADNE_TPX}, where analysis on stopping muons calorimetery using ARIADNE and its agreement to GEANT4 simulation are also detailed. The energy calibration for imaging one of the large 1~$\times$~1~m quadrants, depicted in Figure~\ref{fig:resolution}, was 199.10 
 $\pm$ 1.73 ADU per cm. The energy resolution (preliminary) is approximately {16.73} $\pm$ 0.16\% using tracks of an average length of approx. 22 cm as seen in the dX plot (Figure \ref{fig:resolution}). 

\section{Conclusions}

The ARIADNE$^{+}$ collaboration has successfully demonstrated a dual-phase optical readout with stable detector conditions at a 2~$\times$~2 m active area scale within the CERN Neutrino Platform cold box. This demonstration further substantiates the scalability of the TPX3 camera and THGEM technologies for their application to the kton-scale far LAr detector planned for the DUNE experiment.  

Given the preliminary results presented at NuFACT2022, ARIADNE technology is a serious candidate for the DUNE LAr far detector. Further testing of the technology is required at larger scales, and instrumentation of the NP02 cryostat at the CERN Neutrino Platform with an optical TPC is envisioned, including beam data.   

\vspace{6pt}

\authorcontributions{Conceptualization, K.M. (Konstantinos Mavrokoridis), M.N., B.P., F.P., A.R., F.R., C.T., and J.V.; Funding acquisition, K.M. (Konstantinos Mavrokoridis) and C.T.; Investigation, P.AM., A.D., D.GD., A.J.L., K.M. (Krishanu Majumdar), K.M. (Konstantinos Mavrokoridis), S.R., B.P., A.R. A.SH., and J.V.; Writing, D.GD., A.J.L., K.M. (Krishanu Majumdar), K.M. (Konstantinos Mavrokoridis), B.P., A.R., C.T., and J.V. All authors have read and agreed to the published version of the manuscript.} 

\funding{This research was funded by STFC UKRI Grant No. ST/T007265/1 and ARIADNE ERC Grant No. 677927.} 

\acknowledgments{The authors would like to thank the members of the Mechanical Workshop of the University
of Liverpool’s Physics Department, for their contributions and invaluable expertise. The authors would also like to thank the members of the CERN Neutrino Platform cryogenic team.}

\conflictsofinterest{The authors declare no conflict of interest. The funders had no role in the design of the study; in the collection, analyses, or interpretation of data; in the writing of the manuscript, or in the decision to publish the results.} 



\begin{adjustwidth}{-\extralength}{0cm}
\reftitle{References}

\PublishersNote{}
\end{adjustwidth}

\end{document}